\documentclass[11pt,twoside]{article}
\usepackage{asp2010}

\resetcounters

\begin{document}

\title{Age spread in Galactic star forming region W3 Main}
\author{A. Bik$^1$, Th. Henning,$^1$, A. Stolte,$^2$, W. Brandner$^1$, D. A. Gouliermis$^1$, M. Gennaro$^1$, A. Pasquali$^3$, B. Rochau$^1$, H. Beuther$^1$, N. Ageorges$^4$, W. Seifert$^5$, Y. Wang$^6$, N. Kudryavtseva$^1$}
\affil{$^1$ Max-Planck-Institute for Astronomy, Koenigstuhl 17, 69117, Heidelberg, Germany}
\affil{$^2$ Argelander Institut f\"ur Astronomie, Auf dem H\"ugel 71, 53121 Bonn, Germany}
\affil{$^3$ Astronomisches Rechen Institut, M\"onchhofstrasse 12 - 14, 69120 Heidelberg, Germany}
\affil{$^4$ Max-Planck-Institut f\"ur extraterrestrische Physik, Giessenbachstrasse 1, 85748 Garching, Germany}
\affil{$^5$ Landessternwarte K\"onigstuhl, Zentrum f\"ur Astronomie Heidelberg, K\"onigstuhl 12, 69117 Heidelberg, Germany}
\affil{$^6$ Purple Mountain Observatory, Chinese Academy of Sciences, 210008, Nanjing, PR China}

\begin{abstract}
We present near-infrared JHKs imaging as well as K-band multi-object spectroscopy of the massive stellar content of W3 Main using LUCI at the LBT. We  confirm 13  OB stars by their absorption line spectra in W3 Main and spectral types between O5V and B4V have been found. Three massive Young Stellar Objects are  identified by their emission line spectra and near-infrared excess.   From our spectrophotometric analysis of the massive stars and the nature of their surrounding HII regions we derive the evolutionary sequence of W3 Main and we find  an age spread of  2-3 Myr. 
\end{abstract}

\section{Introduction}

Despite the impact on their surroundings, the formation and early evolution of massive stars is poorly constrained, primarily because of their scarcity and short lifetimes. OB stars form in different environments, ranging from compact, stellar clusters with very high stellar density to associations where the stars are more dispersed over the molecular cloud. This raises the question whether the same star formation mechanism is responsible for the formation of these different structures.  Determining the star formation history, by means of characterizing their stellar content,  will provide insights in how these different regions are formed. Evidence has been found that dense starburst clusters might form in a single burst \citep{Clark05} while OB associations show a large spread in age. 

 To detect the stellar content of these regions directly, near-infrared observations, especially spectroscopy, have proven to be a powerful method to find and characterize the newborn OB stars.  A pure photometric characterization of young stellar clusters is strongly hampered by highly varying extinction, unknown distances and infrared excess of the young cluster members.  A spectroscopic identification, however, results in a unambiguous identification of the (massive) stellar content. Stellar properties, like effective temperature and luminosity, are derived based on the spectral features and extinction, distance and infrared excess can be determined reliably \citep[e.g.][]{Hanson02,Ostarspec05,Puga06,Bik10,Puga10,Gennaro12}. By comparing the effective temperature and  luminosity to main sequence and pre-main-sequence isochrones, the age of a stellar cluster can be derived more reliably  than from photometry alone \citep{Bik10,Bik12}.  
 
We present deep LUCI near-infrared JHKs imaging as well as K-band multi-object-spectroscopy of the massive stellar content of W3 Main. This allows for the first time a spectral classification of the massive stellar content. Using their derived spectral type we discuss the evolutionary status of the massive stars as well as the HII regions in  detail. The different evolutionary phases of the HII regions make W3 Main an ideal target to study age difference in star forming regions.

\section{Near-infrared imaging and spectroscopy}

\begin{figure}[!h]
\includegraphics[width=13cm]{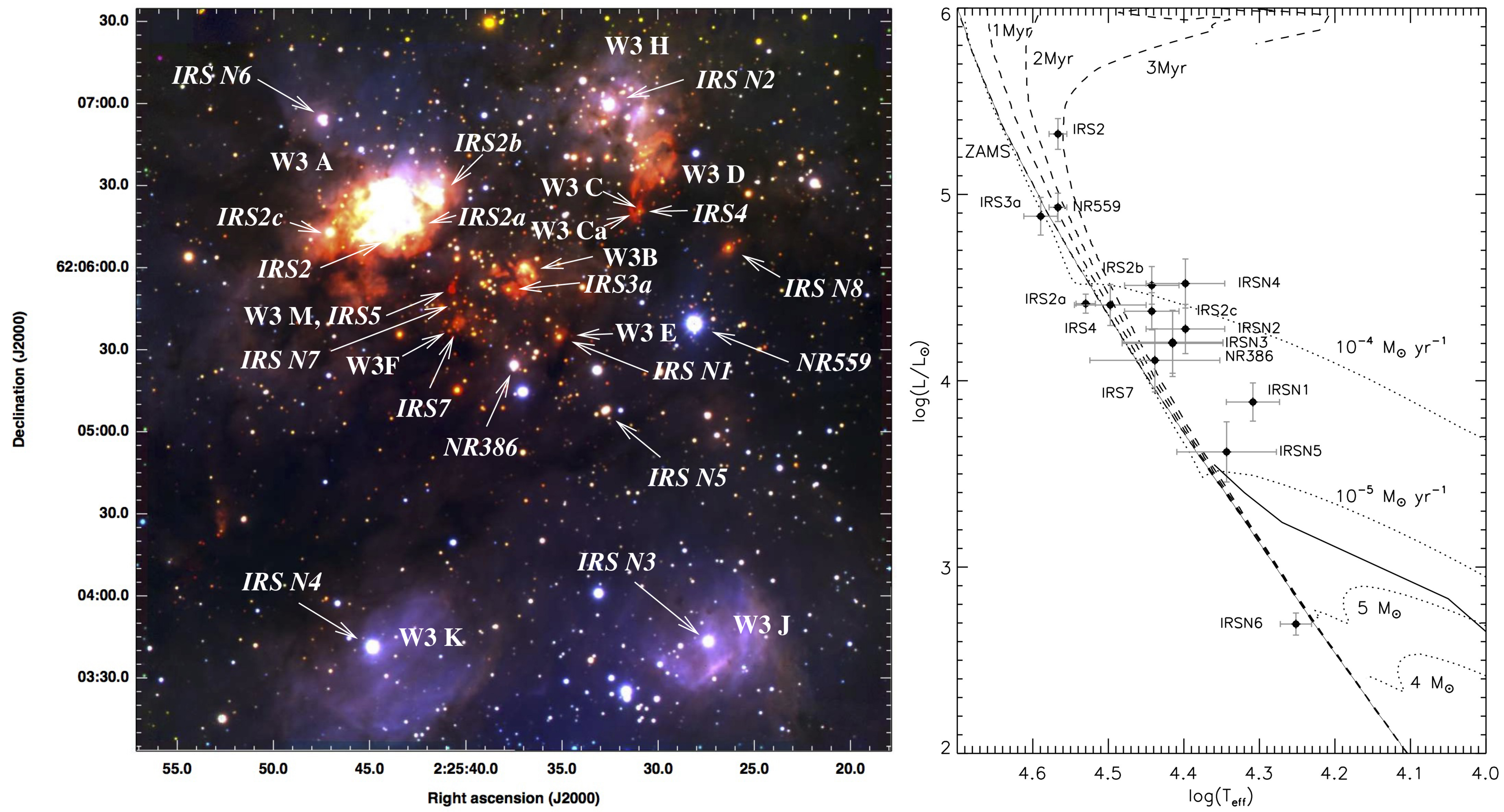}
\caption{\emph{Left:} Near-infrared JHK color composite of W3 Main taken with LUCI at the LBT. \emph{Right:} Hertzsprung Russell Diagram of the massive stars in W3 Main.   Overplotted in the HRD are the main sequence isochrones from the zero age main sequence (ZAMS, solid line) to 3 Myr (dashed lines) from \citet{LeJeune01}, the PMS isochrones for intermediate mass stars  \citep[M $<$ 5 M$_{\odot}$,][dotted lines]{Siess00} as well as the theoretical birth-line (solid line) \citep{Palla90}. 
}\label{fig1}
\end{figure}

JHKs imaging of W3 Main has been obtained with LUCI at the Large Binocular Telescope (LBT) on Mount Graham, Arizona. Fig. \ref{fig1} shows the resulting three color image of a 4' $\times$ 4' field centered on W3 Main.  K-band spectra of  16 candidate massive stars inside W3 Main were obtained with the multi-object spectroscopic mode of LUCI \citep{Bik12}. We classified 13 stars as OB stars from their absorption lines  compared with  K-band spectra of optically visible OB stars \citep{Hanson05,Ostarspec05}.  Three objects are classified as massive Young Stellar Objects: massive stars surrounded by a circumstellar disk. 

After their spectral classification, the bolometric luminosities of the massive stars are calculated and the stars are placed in the Hertzsprung Russell diagram (HRD, Fig. \ref{fig1}) and compared with stellar evolution isochrones.
In the upper regions of the HRD (log L/L$_{\odot}$ $\geq$ 5), the main sequence isochrones indicate that the most massive stars already show significant evolution  after a few Myr. Two stars are located in the upper regions of the HRD, IRS2 and IRS3a.  IRS2 is located to the right of the ZAMS and its location is more consistent with the 2-3 Myr isochrones. The foreground extinction is not very extreme, therefore the  location of IRS2 in the HRD does not vary a lot by changing the extinction law. IRS3a (O5V -- O7V), however, is very reddened (A$_{Ks}$=5.42 $\pm$ 0.79 mag) and its location in the HRD depends very strongly on the adopted extinction law and its age cannot be constrained accurately.

\section{Age spread}

W3 Main harbors several different evolutionary stages of HII regions, ranging from very young hyper-compact HII (HCHII) regions (few 10$^3$ years), ultra-compact HII (UCHII) regions \citep[$\sim 10^5$ years,][]{WoodIRAS89} to evolved,  diffuse HII regions (few 10$^6$ years). All these regions are most likely formed out of the  same molecular cloud. This provides the possibility to study the evolution of young HII regions and their stellar content in great detail.  \citet{Tieftrunk97} derived an evolutionary sequence for the HII regions in W3 Main, based on the morphology of the radio sources.

\begin{figure}[!ht]
\includegraphics[width=13cm]{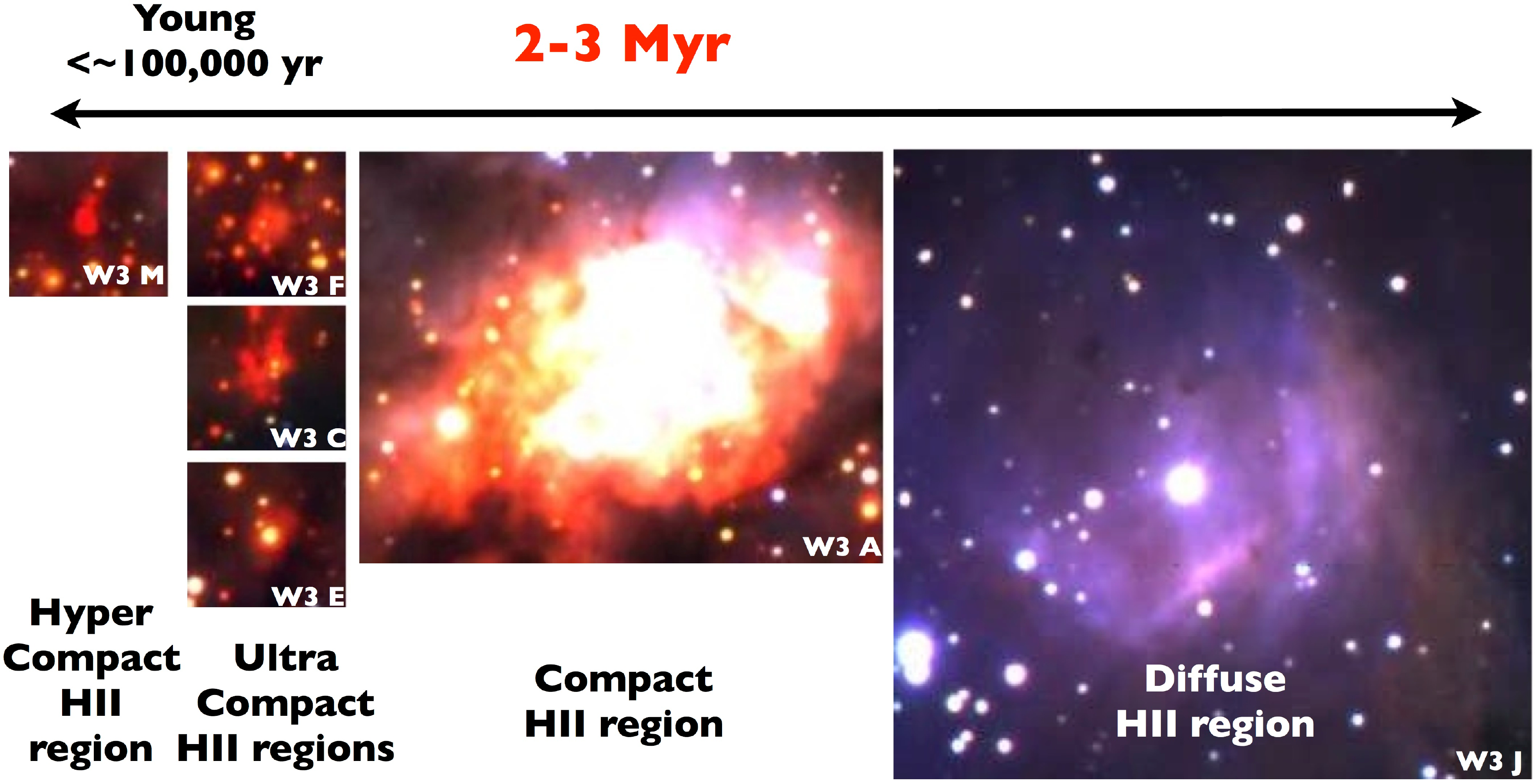}
\caption{Near-infrared 3 color composites of the different kinds of HII regions in W3 Main ordered from young to the more evolved regions. The age of W3A is the age derived for IRS2, the most massive star in W3 Main.}\label{fig2}
\end{figure}

The youngest are the HCHII regions W3 M and W3 Ca, with the UCHII regions W3 F, W3 C and W3 E, slightly older, the compact HII regions W3 B and W3 A even more evolved, and the diffuse HII regions, W3 K and W3 J being the oldest HII regions in W3 Main (Fig. \ref{fig2}).  Additionally, evidence has been found that W3 A is interacting with the youngest sites of star formation and could have triggered the formation of IRS5 \citep{Wang12}. 

This evolutionary sequence can be compared to the ages of the massive stars deduced from their position in the HRD.
We have detected OB stars in three diffuse HII regions, two compact HII regions and three UCHII regions. The HCHII region W3 M harbors the high-mass protostar IRS5.  Additionally three stars have no detectable HII region associated with them (IRS N5 and the massive YSOs IRS N7 and IRS N8).  The position of IRS2 in the HRD suggests an age of  2-3 Myr, consistent with its location in a relatively evolved compact HII region (W3 A).

\section{Discussion}

Based on the presence of different evolutionary stages of HII regions as well as the location of the most massive stars in the HRD, we can conclude that an age spread of  2-3 Myr is most likely present for the massive stars in W3 Main. A growing number of  young stellar clusters show evidence for an age spread, usually based on the analysis of their PMS population in the HRD. In Orion an age spread of a few Myr has been found  \citep{Reggiani11} as well as in LH95 in the LMC \citep{DaRio10}. In starburst clusters, however, no age spread has been found for Westerlund 1 and NGC3603 \citep{Clark05,Kudryavtseva12} based on both the massive stars and the lower-mass PMS stars. This suggests that W3 Main is not formed in one star formation burst as expected for starburst cluster, but, more likely, through a temporal sequence of star formation events.



\end{document}